\documentclass[12pt,a4paper]{article}
\usepackage[british]{babel}
\usepackage{epsfig}

\begin{document}
\date{}

\title{
{\vspace{-20mm} \normalsize
\hfill \parbox[t]{50mm}{DESY 03-019  \\
                        MS-TP-03-03}}\\[20mm]
        Partially quenched chiral perturbation theory       \\
        and numerical simulations}

\author{qq+q Collaboration                                  \\[0.5em]
        F. Farchioni                                        \\[0.5em]
        Westf\"alische Wilhelms-Universit\"at M\"unster,    \\
        Institut f\"ur Theoretische Physik,                 \\
        Wilhelm-Klemm-Strasse 9, D-48149 M\"unster, Germany \\[1em]
        C. Gebert, I. Montvay, E. Scholz, L. Scorzato       \\[0.5em]
        Deutsches Elektronen-Synchrotron DESY               \\
        Notkestr.\,85, D-22603 Hamburg, Germany}

\newcommand{\be}{\begin{equation}}                                              
\newcommand{\ee}{\end{equation}}                                                
\newcommand{\half}{\frac{1}{2}}                                                 
\newcommand{\rar}{\rightarrow}                                                  
\newcommand{\lar}{\leftarrow}
\newcommand{\LCB}{\raisebox{-0.3ex}{\mbox{\LARGE$\left\{\right.$}}}
\newcommand{\RCB}{\raisebox{-0.3ex}{\mbox{\LARGE$\left.\right\}$}}}
\newcommand{\U}{\mathrm{U}}
\newcommand{\SU}{\mathrm{SU}}

\maketitle

\abstract{
 The dependence of the pseudoscalar meson mass and decay constant is
 compared to one-loop Partially Quenched Chiral Perturbation Theory
 (PQChPT) in a numerical simulation with two light dynamical quarks.
 The characteristic behaviour with chiral logarithms is observed.
 The values of the fitted PQChPT-parameters are in a range close to the
 expectation in continuum in spite of the fact that the lattice spacing
 is still large, namely $a\simeq 0.28\, {\rm fm}$.}

\section{Introduction}\label{sec1}
 In numerical Monte Carlo simulations of QCD Chiral Perturbation
 Theory (ChPT) \cite{CHPT} is often used to guide the extrapolation to
 the physical values of the three light quark masses ($m_u \simeq m_d$
 and $m_s$).
 In this procedure not only the lattice gauge theory results are
 established but also useful information is obtained about the values of
 the {\em Gasser-Leutwyler parameters} of ChPT.
 In fact, recently several groups explored this possibility in
 quenched \cite{ALPHA:CHPT} and unquenched simulations with Wilson-type
 \cite{UKQCD:CHPT} and staggered \cite{STAGGERED1,STAGGERED2} quarks.
 (For a review see \cite{WITTIG}.)

 In order to achieve small systematic errors the simulations themselves
 have to be performed in a range of quark masses where the applied
 one-loop (NLO) ChPT-formulas give a good approximation.
 In particular, the characteristic {\em chiral logarithms} have to
 appear in the quark mass dependence of different physical quantities.
 This applies both to original ChPT as well as to PQChPT
 \cite{SHARPE,GOLT-LEUNG}.
 However, in most recent simulations -- especially with Wilson-type
 quarks -- this condition is not fulfilled because they are performed
 in the range $m_{u,d} \geq \half m_s$.
 Estimates based on present knowledge of the ChPT parameters indicate
 (see, for instance, \cite{SHARPE-SHORESH}) that at least 
 $m_{u,d} \leq \frac{1}{4}m_s - \frac{1}{5}m_s$ has to be reached.
 (See also the summary of the panel discussion at the Boston Lattice
 Conference \cite{PANEL}.)
 Another requirement is that the virtual effects of the $s$-quark
 also has to be taken into account by simulating with three light
 dynamical quarks.

 Since the dynamical quarks in most unquenched simulations do not
 satisfy the above bound, it is not surprising that the chiral
 logarithms have not been observed 
 \cite{CPPACS,UKQCD,JLQCD:CHLOGM,JLQCD:CHLOGF,CPPACS:Namekawa,JLQCD}.
 This was the main motivation of our collaboration to start exploring
 the possibility of simulating QCD with light quarks in the range
 $m_{ud} \leq \half m_s$ \cite{NF2TEST,BOSTON,PRICE,GEBERT,HEIDEL}.
 In these simulations we use the {\em two-step multi-boson} (TSMB)
 algorithm for dynamical fermions \cite{TSMB} and consider for the
 moment $N_s=2$ dynamical ``sea'' quarks.
 The case of $N_s=3$ is also under study \cite{BERLIN}.
 Our first simulations were oriented towards the investigation of
 simulation costs as a function of the quark mass and were performed
 on modest size lattices (typically $8^3 \cdot 16$) with lattice
 spacings of the order $a\simeq 0.27\, {\rm fm}$ -- a value where
 continuum behaviour is not necessarily expected.
 Therefore, it came to us as a surprise that plotting the pseudoscalar
 (``pion'') mass and decay constant as a function of the quark mass
 (in the form suggested by \cite{LEUTWYLER,DURR}) the chiral
 logarithmic behaviour has been qualitatively displayed
 \cite{PRICE,GEBERT}.

 Encouraged by this result we picked out a point with
 $m_{ud} \simeq \frac{1}{4}m_s$ and performed a high statistics run on
 $16^4$ lattice in order to study the dependence on the valence
 quark mass in a sufficiently large physical volume.
 This has the advantage that by taking ratios of the masses and
 decay constants the $Z$-factors of renormalization cancel.
 This removes an uncertainty in \cite{PRICE,GEBERT} where the
 $Z$-factors have been neglected by setting them to $Z \equiv 1$.
 In our analysis of simulation data we applied PQChPT for Wilson
 lattice fermions \cite{RUPAK-SHORESH} which take into account leading
 lattice artefacts of ${\cal O}(a)$.

 The plan of this paper is as follows: in the next section the one-loop
 PQChPT formulas for Wilson lattice fermions will be recapitulated.
 In section 3 the numerical simulation data will be analyzed and
 discussed.

\section{PQChPT formulas}\label{sec2}
 Our analysis of the valence quark dependence of the pseudoscalar
 mass ($m_\pi$) and decay constant ($f_\pi$) is based on the one-loop
 PQChPT formulas for the Wilson lattice action as derived in
 \cite{RUPAK-SHORESH}.
 Instead of the quantities with dimension mass-square $\chi_A$ and
 $\rho_A$ of ref.~\cite{RUPAK-SHORESH} we prefer to use the
 dimensionless ones
\be \label{eq01}
\chi_A \equiv \frac{2B_0 m_q}{f_0^2} \ , \hspace{3em}
\rho_A \equiv \frac{2W_0 ac_{SW}}{f_0^2} \ .
\ee
 Here $m_q$ is the quark mass, $a$ the lattice spacing, $B_0$ and $W_0$
 are parameters of dimension mass and (mass)$^3$, respectively, which
 appear in the leading order (LO) chiral effective Lagrangian, $c_{SW}$
 is the coefficient of the ${\cal O}(a)$ chiral symmetry breaking term
 and $f_0$ is the value of the pion decay constant at zero quark mass.
 (Its normalization is such that the physical value is
 $f_0 \simeq 93\, {\rm MeV}$.)
 In ref.~\cite{RUPAK-SHORESH} the case of three non-degenerate quark
 flavours is considered.
 Here we consider a general number $N_s$ of equal mass sea quarks.

 The next to leading order (NLO) PQChPT formula for the pion decay
 constant is in this case:
\begin{eqnarray} \frac{f_{AB}}{f_0} \hspace*{-1.4em} && =
1 - \frac{N_s}{128\pi^2} \LCB (\chi_A+\chi_S+\rho_A+\rho_S)
\log\left(\half(\chi_A+\chi_S+\rho_A+\rho_S)\right)
\nonumber
\\[0.5em] &&
+ (\chi_B+\chi_S+\rho_B+\rho_S)
\log\left(\half(\chi_B+\chi_S+\rho_B+\rho_S)\right) \RCB
\nonumber
\\ &&
+ \frac{1}{64N_s\pi^2} \LCB 
\chi_A+\chi_B+\rho_A+\rho_B-2\chi_S-2\rho_S + 
(\chi_B-\chi_A+\rho_B-\rho_A)^{-1}
\nonumber
\\[0.5em] &&
\cdot \left[
2(\chi_A+\rho_A)(\chi_B+\rho_B) - (\chi_S+\rho_S)
(\chi_A+\chi_B+\rho_A+\rho_B) 
\log\left(\frac{\chi_A+\rho_A}{\chi_B+\rho_B}\right) \right] \RCB
\nonumber
\\[1.0em] && \label{eq02}
+ 2\bar{L}_5 (\chi_A+\chi_B) + 2\bar{W}_5 (\rho_A+\rho_B)
+ 4N_s\bar{L}_4\chi_S + 4N_s\bar{W}_4\rho_S \ .
\end{eqnarray}
 Here A and B denote generic quark indices: S will be the label for
 the sea quarks V for valence quarks.
 For the pion mass squared we have:
\begin{eqnarray} \frac{m_{AB}^2}{f_0^2} \hspace*{-1.4em} && =
\half (\chi_A+\chi_B+\rho_A+\rho_B)
+ \frac{1}{32N_s\pi^2}\,
  \frac{(\chi_A+\chi_B+\rho_A+\rho_B)}{(\chi_B-\chi_A+\rho_B-\rho_A)}
\nonumber
\\[0.5em] &&
\cdot \LCB 
  (\chi_A+\rho_A)(\chi_S-\chi_A+\rho_S-\rho_A) \log(\chi_A+\rho_A)
\nonumber
\\[0.5em] &&
- (\chi_B+\rho_B)(\chi_S-\chi_B+\rho_S-\rho_B) \log(\chi_B+\rho_A) 
\RCB
\nonumber
\\[1.0em] && \label{eq03}
+ 4N_s(2\bar{L}_6-\bar{L}_4)\chi_S(\chi_A+\chi_B)
+ 2\,(2\bar{L}_8-\bar{L}_5)(\chi_A+\chi_B)^2
\nonumber
\\[0.5em] &&
+ 4N_s(\bar{W}_6-\bar{L}_4)\chi_S(\rho_A+\rho_B)
+ 4N_s(\bar{W}_6-\bar{W}_4)\rho_S(\chi_A+\chi_B)
\nonumber
\\[0.5em] &&
+ 2\,(2\bar{W}_8-\bar{W}_5-\bar{L}_5)(\chi_A+\chi_B)(\rho_A+\rho_B) \ .
\end{eqnarray}
 The NLO parameters $\bar{L}_k$ and $\bar{W}_k$ are related to $L_k$
 and $W_k$ in ref.~\cite{SHARPE-SHORESH,RUPAK-SHORESH} by
\be \label{eq04}
\bar{L}_k \equiv L_k - c_k \log(f_0^2) \ , \hspace{3em}
\bar{W}_k \equiv W_k - d_k \log(f_0^2) \ ,
\ee
 where the coefficients of the logarithms are given by
\be \label{eq05}
c_4 = \frac{1}{256\pi^2} \ , \hspace{1.5em}
c_5 = \frac{N_s}{256\pi^2} \ , \hspace{1.5em}
c_6 = \frac{(N_s^2+2)}{512N_s^2\pi^2} \ , \hspace{1.5em}
c_8 = \frac{(N_s^2-4)}{512N_s\pi^2} \ ,
\ee
 respectively,
\be \label{eq06}
d_4 = \frac{1}{256\pi^2} \ , \hspace{1.5em}
d_5 = \frac{N_s}{256\pi^2} \ , \hspace{1.5em}
d_6 = \frac{(N_s^2+2)}{256N_s^2\pi^2} \ , \hspace{1.5em}
d_8 = \frac{(N_s^2-4)}{256N_s\pi^2} \ .
\ee
 The relations in (\ref{eq04}) have the unpleasant  feature that
 logarithms of a dimensionful quantity appear.
 One can avoid this by introducing
\be \label{eq07}
L_k^\prime \equiv L_k - c_k \log(\mu^2) \ , \hspace{3em}
W_k^\prime \equiv W_k - d_k \log(\mu^2) \ ,
\ee
 where $\mu$ is the mass scale introduced by dimensional regularization.
 Since $L_k$ and $W_k$ depend on $\mu$ the choice of it in the
 logarithm is natural.
 In terms of $L_k^\prime$ and $W_k^\prime$ we have
\be \label{eq08}
\bar{L}_k = L_k^\prime - 
c_k \log\left(\frac{f_0^2}{\mu^2}\right) \ , \hspace{3em}
\bar{W}_k = W_k^\prime - 
d_k \log\left(\frac{f_0^2}{\mu^2}\right) \ .
\ee

 Note that the NLO parameters $\alpha_k$ in ref.~\cite{SHARPE} are
 related to $L_k^\prime$ by
\be \label{eq09}
\alpha_k = 128 \pi^2 L_k^\prime \ .
\ee
 The {\em universal low energy scales} $\Lambda_{3,4}$ in
 ref.~\cite{LEUTWYLER,DURR} can be expressed, in the case of $N_s=2$,
 by the following combinations of the coefficients $\bar{L}_k$:
\begin{eqnarray}
-\frac{1}{256\pi^2}\log\frac{\Lambda_3^2}{f_0^2} \hspace*{-1.0em}&&
=\; 2\bar{L}_8-\bar{L}_5+4\bar{L}_6-2\bar{L}_4 \ ,
\nonumber
\\[0.5em] \label{eq10}
\frac{1}{64\pi^2}\log\frac{\Lambda_4^2}{f_0^2} \hspace*{-1.0em}&&
=\; 2\bar{L}_4+\bar{L}_5 \ .
\end{eqnarray}

 In this paper we are interested in the valence quark mass dependence
 of $f_\pi$ and $m_\pi^2$ for fixed sea quark mass parameter $\chi_S$.
 Therefore it is natural to introduce the ratios of the other mass
 parameters to $\chi_S$:
\be \label{eq11}
\xi   \equiv \frac{\chi_V}{\chi_S} \ , \hspace{3em}
\eta  \equiv \frac{\rho_S}{\chi_S} \ , \hspace{3em}
\zeta \equiv \frac{\rho_V}{\rho_S} =  \frac{\rho_V}{\eta\chi_S} \ .
\ee
 Once relations (\ref{eq11}) are substituted in
 (\ref{eq02})-(\ref{eq03}), the logarithmic dependence on $\chi_S$ can
 be absorbed in the NLO parameters if one introduces
\begin{eqnarray}
L_{Sk} \hspace*{-1.0em} &&\equiv \hspace*{0.7em} 
\bar{L}_k - c_k\log(\chi_S) 
\;=\; L_k^\prime - c_k\log\left(\frac{f_0^2}{\mu^2}\chi_S\right) \ ,
\nonumber
\\[0.5em] \label{eq12}
W_{Sk} \hspace*{-1.0em} &&\equiv \hspace*{0.7em} 
\bar{W}_k - d_k\log(\chi_S) 
\;=\; W_k^\prime - d_k\log\left(\frac{f_0^2}{\mu^2}\chi_S\right) \ .
\end{eqnarray}
 Let us note that the argument of the last logarithms here can also be
 written as
\be \label{eq13}
\frac{f_0^2}{\mu^2}\,\chi_S \,=\, \frac{2B_0m_{qS}}{\mu^2} \ .
\ee

 In this paper we keep the sea quark mass ($\chi_S$) fixed and
 vary the valence quark mass ($\chi_V=\xi\chi_S$).
 Expanding the ratio of decay constants up to first order in the
 one-loop corrections one obtains
\begin{eqnarray}
Rf_{VV} \hspace*{-1.5em}&& \equiv \frac{f_{VV}}{f_{SS}}
= 1 + 4(\xi-1)\chi_S L_{S5}
\nonumber
\\[0.5em] && \label{eq14}
- \frac{N_s\chi_S}{64\pi^2}(1+\xi+2\eta)\log\frac{1+\xi+2\eta}{2}
+ \frac{N_s\chi_S}{32\pi^2}(1+\eta)\log(1+\eta) \ ,
\end{eqnarray}
 and similarly
\begin{eqnarray}
Rf_{VS} \hspace*{-1.5em}&& \equiv \frac{f_{VS}}{f_{SS}}
= 1 + 2(\xi-1)\chi_S L_{S5} + \frac{\chi_S}{64N_s\pi^2}(\xi-1)
- \frac{\chi_S}{64N_s\pi^2}(1+\eta)\log\frac{\xi+\eta}{1+\eta}
\nonumber
\\[0.5em] && \label{eq15}
- \frac{N_s\chi_S}{128\pi^2}(1+\xi+2\eta)\log\frac{1+\xi+2\eta}{2}
+ \frac{N_s\chi_S}{64\pi^2}(1+\eta)\log(1+\eta) \ .
\end{eqnarray}
 In case of the ratios of $m_\pi^2$ we expand up to first order in the
 ``correction'' which now also includes the ${\cal O}(a)$ terms of
 the tree-level expressions:
\begin{eqnarray}
Rm_{VV} \hspace*{-1.5em}&& \equiv \frac{m_{VV}^2}{m_{SS}^2}
= \xi+\eta-\eta\xi
\nonumber
\\[0.5em] &&
+ 8\xi(\xi-1)\chi_S(2L_{S8}-L_{S5})
+ 8N_s(\xi-1)\eta\chi_S(L_{S4}-W_{S6})
\nonumber
\\[0.5em] &&
+ \frac{\chi_S}{16N_s\pi^2}(\xi-1)(\xi+\eta)
- \frac{\chi_S}{16N_s\pi^2}\xi(1+2\eta)\log(1+\eta)
\nonumber
\\[0.5em] && \label{eq16}
+ \frac{\chi_S}{16N_s\pi^2}(2\xi^2-\xi-\eta+3\eta\xi)\log(\xi+\eta) \ ,
\end{eqnarray}
 and
\begin{eqnarray}
Rm_{VS} \hspace*{-1.5em}&& \equiv \frac{m_{VS}^2}{m_{SS}^2}
= \half(1+\xi+\eta-\eta\xi)
\nonumber
\\[0.5em] &&
+ 2(\xi+1)(\xi-1)\chi_S(2L_{S8}-L_{S5})
+ 4N_s(\xi-1)\eta\chi_S(L_{S4}-W_{S6})
\nonumber
\\[0.5em] &&
- \frac{\chi_S}{32N_s\pi^2}(\xi+1)(1+2\eta)\log(1+\eta)
\nonumber
\\[0.5em] && \label{eq17}
+ \frac{\chi_S}{32N_s\pi^2}(\xi^2+\xi+\eta+3\eta\xi)\log(\xi+\eta) \ .
\end{eqnarray}
 In these expressions it is assumed that the ${\cal O}(a)$ mass terms
 $\rho$ are the same for valence quarks and the sea quark, namely
 $\rho_V=\rho_S$.
 This is the case if only the hopping parameter $\kappa$ is changed.
 Changing $\rho_V=\zeta\eta\chi_S$ can be investigated by changing
 the Wilson-parameter $r$ in the Wilson fermion action, too.

 Up to now we considered the valence quark mass dependence for
 unchanged sea quark masses.
 Let us remark that the formulas for the sea quark mass dependence can
 also be written in a similar form as eqs.~(\ref{eq14})-(\ref{eq17}).
 In this case it is advantageous to fix a {\em reference sea quark mass}
 $\chi_R$ (see \cite{ALPHA:CHPT}) and introduce the variables
\be \label{eq18}
\sigma   \equiv \frac{\chi_S}{\chi_R} \ , \hspace{3em}
\omega  \equiv \frac{\rho_R}{\chi_R} \ , \hspace{3em}
\tau \equiv \frac{\rho_S}{\rho_R} =  \frac{\rho_S}{\omega\chi_R} \ .
\ee
 Instead of the NLO parameters in (\ref{eq12}) the appropriate ones
 are then obviously
\be \label{eq19}
L_{Rk} \equiv L_k^\prime - 
c_k\log\left( \frac{f_0^2}{\mu^2}\chi_R \right) \ , \hspace{2em}
W_{Rk} \equiv W_k^\prime - 
d_k\log\left( \frac{f_0^2}{\mu^2}\chi_R \right) \ .
\ee
%

\section{Numerical results}\label{sec3}
 We performed Monte Carlo simulations with $N_s=2$ degenerate sea quarks
 on a $16^4$ lattice at $\beta=4.68$, $\kappa=0.195$ and investigated
 the valence quark mass dependence at
 $\kappa=0.1955,\,0.1945,\,0.1940,\,0.1930,\,0.1920$.
 The statistics corresponds to 1180 gauge field configurations.
 The error analysis was based on the {\em linearization method}
 \cite{ALPHA:BENCHMARK}.
 Since $r_0/a = 1.76(6)$ the lattice spacing is
 $a \simeq 0.28\, {\rm fm}$.
 This means that the physical lattice extension is rather large:
 $L \simeq 4.5\, {\rm fm}$.
 The value of the quark mass parameter is given by $am_\pi=0.519(1)$ as
 $M_r \equiv (r_0 m_\pi)^2 \simeq 0.83$.
 This is about $\frac{1}{4}$ of the value of $M_r$ for the strange
 quark $M_r^{strange} \simeq 3.1$.

 The ratios in eqs.~(\ref{eq14})-(\ref{eq17}) as a function of the
 quark mass ratio ($\xi$) depend on five parameters, namely
 $\chi_S,\,\eta,\,L_{S5},(2L_{S8}-L_{S5})$ and $(L_{S4}-W_{S6})$.
 With our choice of the valence hopping parameters and with our
 statistics most of the multi-parameter fits were unstable therefore
 our analysis is based one a sequence of single or double parameter fits.
 The stability of the multi-parameter fits can be improved by optimizing
 the choice of valence quark mass values, which we did not exploit this
 time.
\begin{figure}[ht]
\vspace*{-1mm}
\begin{center}
\epsfig{file=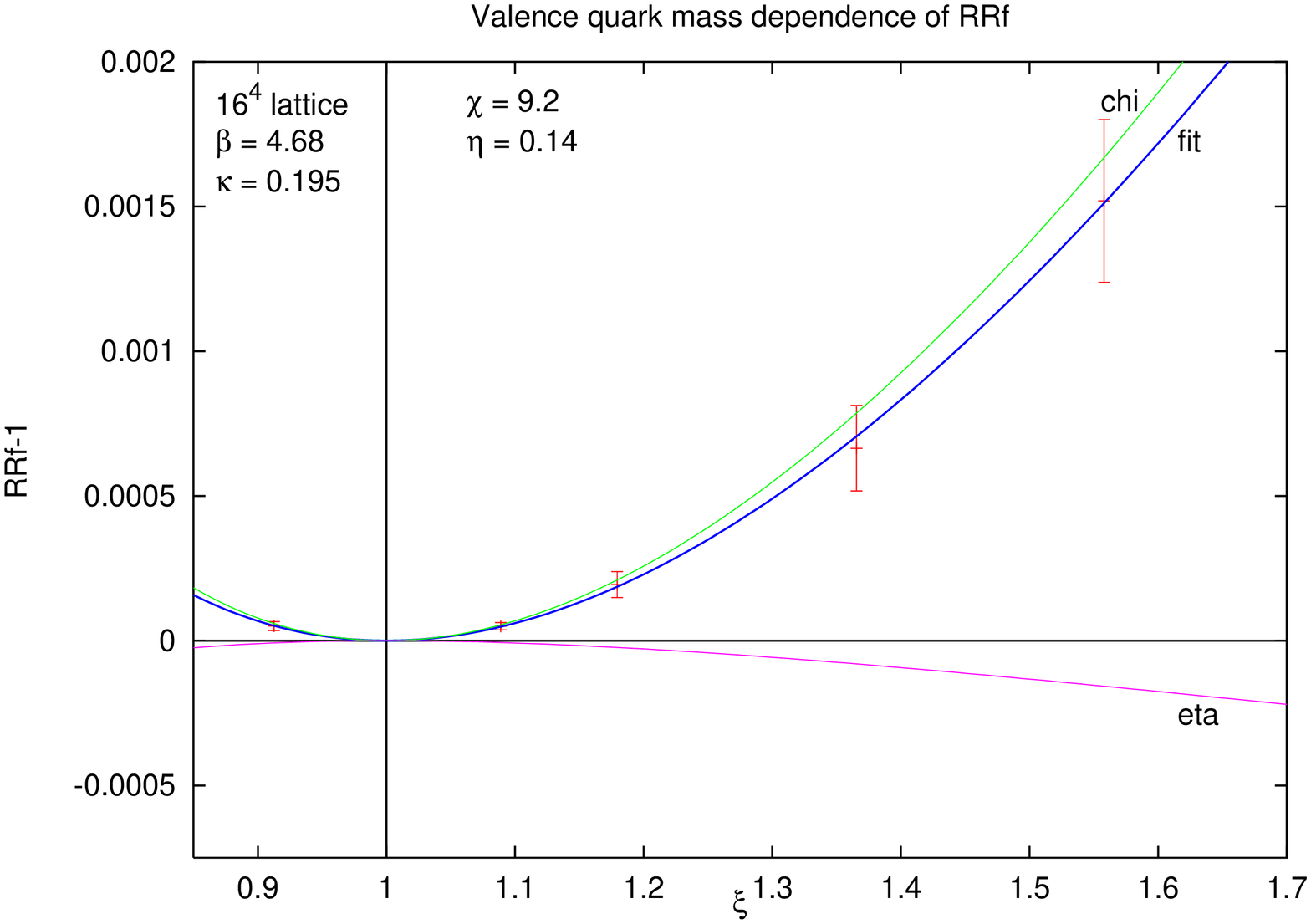,
        width=100mm,height=140mm,angle=-90,
        bbllx=50pt,bblly=30pt,bburx=554pt,bbury=800pt}
\vspace*{-1mm}
\parbox{12cm}{\em\caption{\label{fig01}
 The valence quark mass dependence of the double ratio of pion decay
 constants $RRf$.
 Besides the ``fit'' the two other curves show the ${\cal O}(a)$
 contribution (``eta'') and the physical contribution obtained at
 $\eta=0$ (``chi'').}}
\end{center}
\vspace*{-1mm}
\end{figure}
\begin{figure}[ht]
\vspace*{-1mm}
\begin{center}
\epsfig{file=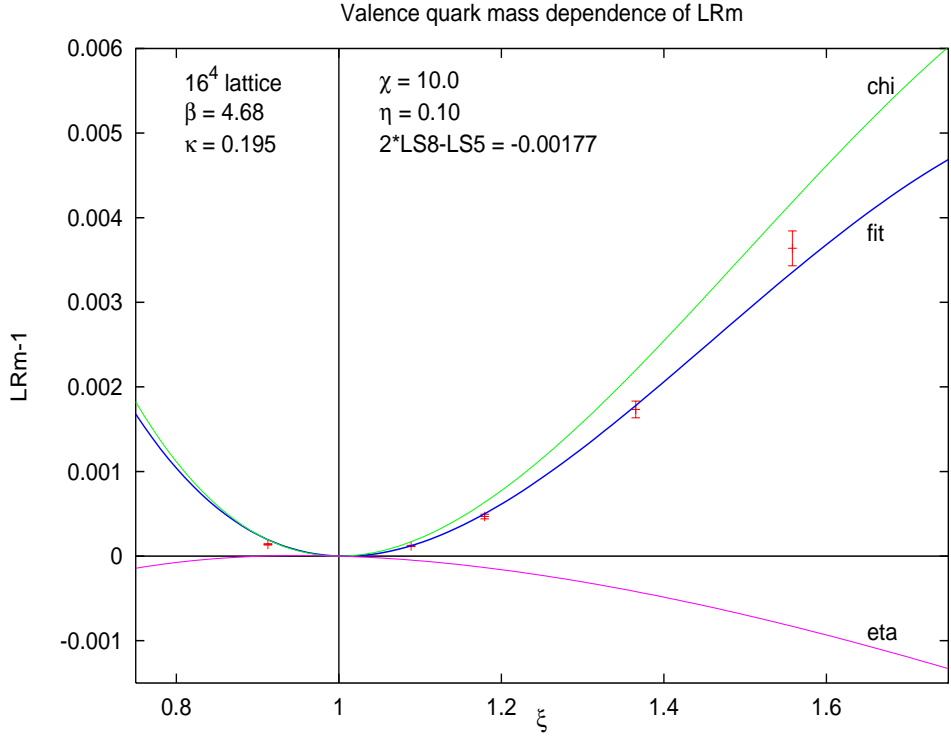,
        width=100mm,height=140mm,angle=-90,
        bbllx=50pt,bblly=30pt,bburx=554pt,bbury=800pt}
\vspace*{-1mm}
\parbox{12cm}{\em\caption{\label{fig02}
 The valence quark mass dependence of the linear combination of pion
 mass-squared ratios $LRm$.
 Besides the ``fit'' the two other curves show the ${\cal O}(a)$
 contribution (``eta'') and the physical contribution obtained at
 $\eta=0$ (``chi'').}}
\end{center}
\vspace*{-1mm}
\end{figure}
\begin{figure}[ht]
\vspace*{-1mm}
\begin{center}
\epsfig{file=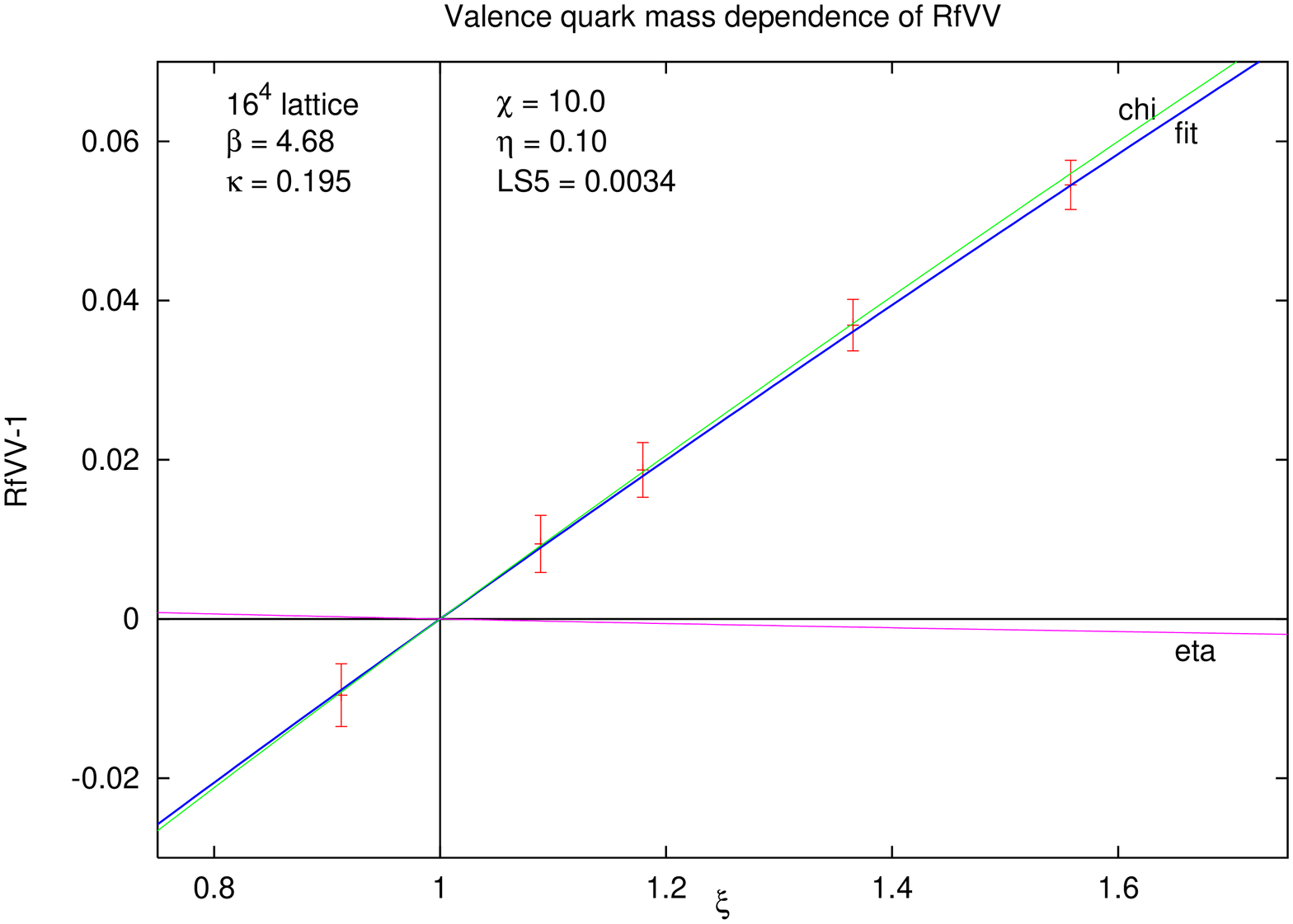,
        width=100mm,height=140mm,angle=-90,
        bbllx=50pt,bblly=30pt,bburx=554pt,bbury=800pt}
\vspace*{-1mm}
\parbox{12cm}{\em\caption{\label{fig03}
 The valence quark mass dependence of the ratio of pion decay constants
 $Rf_{VV}$.
 Besides the ``fit'' the two other curves show the ${\cal O}(a)$
 contribution (``eta'') and the physical contribution obtained at
 $\eta=0$ (``chi'').}}
\end{center}
\vspace*{-1mm}
\end{figure}

 A very useful quantity is the double ratio of decay constants
 \cite{JLQCD:CHLOGF} which does not depend on any of the NLO
 coefficients.
 In other words there one can see the chiral logarithms alone.
 The NLO formula is:
\be \label{eq20}
RRf \equiv \frac{f_{VS}^2}{f_{VV}f_{SS}} =
1 + \frac{\chi_S}{32N_s\pi^2}(\xi-1)
- \frac{\chi_S}{32N_s\pi^2}(1+\eta)\log\frac{\xi+\eta}{1+\eta} \ .
\ee
 Because of the ${\cal O}(a)$ contributions this has two parameters:
 $\chi_S$ and $\eta=\rho_S/\chi_S$.
 For performing two-parameter fits some timeslice distance pairs were
 chosen and the fits of the pion- and quark mass were taken from them.
 Useful choices are, for instance, 4-5, 4-6 or 5-6.
 (The way physical quantities were obtained has been described in detail
 in \cite{NF2TEST}).
 The result of the two-parameter fit was (see figure \ref{fig01}):
\be \label{eq21}
\chi_S = 9.2 \pm 2.5 \ , \hspace{3em}
\eta = 0.14 \pm 0.30 \ .
\ee
 The value of $\chi_S$ has the right order of magnitude.
 Indeed, from the axial Ward identity we obtain
 $r_0m_{qS}=0.06\,Z_q^{-1}$.
 Using this and the phenomenological estimates \cite{DURR}
 $r_0f_0=0.23,\,r_0B_0=7.0$, where the value of $B_0$ refers to the
 $\overline{MS}$ scheme at $\mu=2\,{\rm GeV}$, we deduce
 $\chi_S^{estimate} \simeq 16\,Z_q^{-1}$.
 Here $Z_q$ is an unknown $Z$-factor relating the bare lattice quark
 mass to the renormalized one at 2~GeV, which is typically of
 ${\cal O}(1)$.
 Another estimate can be obtained by using the tree-level ChPT
 formula $\chi_S^{estimate} \approx M_r/(r_0 f_0)^2 \simeq 15.7$.

 As one can see from (\ref{eq21}), the value of the ratio
 $\eta = \rho_S/\chi_S$ has a large error in the fit.
 The reason is that the functional form of the terms due to the
 leading ${\cal O}(a)$ lattice artefacts is not very different from
 those coming from the non-zero quark mass.
 Nevertheless, they are different and an optimized choice of
 $\xi$-values plus high statistics allowing for stable multi-parameter
 fits makes the distinction of the two sorts of chiral symmetry
 breaking possible.
 In the rest of this paper we only consider one- and two-parameter
 fits with fixed $\eta$ in the range given by (\ref{eq21}).

 After determining $\chi_S$ and $\eta=\rho_S/\chi_S$ from the double
 ratio $RRf$ one can fit the other ratios to obtain estimates of the
 NLO coefficients.
 A nice linear combination of mass-squared ratios is:
\begin{eqnarray}
LRm \hspace*{-1.5em}&& \equiv 2Rm_{VS}-Rm_{VV}
= 1 - 4(\xi-1)^2 \chi_S(2L_{S8}-L_{S5})
- \frac{\chi_S}{16N_s\pi^2}(\xi-1)(\xi+\eta)
\nonumber
\\[0.5em] && \label{eq22}
- \frac{\chi_S}{16N_s\pi^2}(1+2\eta)\log(1+\eta)
- \frac{\chi_S}{16N_s\pi^2}(\xi^2-2\xi-2\eta)\log(\xi+\eta) \ .
\end{eqnarray}
 This has only a single new parameter $(2L_{S8}-L_{S5})$ and the
 statistical errors are small, therefore one can also perform a
 two-parameter fit of $\chi_S$ and $(2L_{S8}-L_{S5})$ with the result
 ($\eta = 0.05$ fixed, errors in last digits given in parentheses):
\be \label{eq23}
2L_{S8}-L_{S5} = -0.00203(5) \ , \hspace{2em}
2\alpha_8-\alpha_5 = 0.85(6) \ , \hspace{2em}
\chi_S = 5.2(1.1) \ .
\ee
 $\alpha_k$ denote the NLO parameters in (\ref{eq09}) taken at the
 renormalization scale $\mu=4\pi f_0$.
 Fixing both $\chi_S = 10.0$ and $\eta = 0.10$ gives:
\be \label{eq24}
2L_{S8}-L_{S5} = -0.00177(3) \ , \hspace{2em}
2\alpha_8-\alpha_5 = 0.58(3) \ .
\ee
 As figure \ref{fig02} shows, with these parameters the last point
 is not perfectly fitted.
 A perfect fit is obtained with the parameters in (\ref{eq23}).

 The value of $L_{S5}$ can be determined from $Rf_{VV}$.
 ($Rf_{VS}$ gives very similar results.)
 In this case the statistical errors are larger, therefore only a
 single parameter fit is useful.
 The result for fixed $\chi_S = 10.0$ and $\eta = 0.10$ is (see figure
 \ref{fig03}):
\be \label{eq25}
L_{S5} = 0.0034(1) \ , \hspace{3em}
\alpha_5 = 1.55(24) \ .
\ee
 The errors given in (\ref{eq23})-(\ref{eq25}) are the ones for the
 specified values of fixed parameters.
 The overall error is, of course, larger --  as one can see, for
 instance, by comparing (\ref{eq23}) and (\ref{eq24}).
 The values of $(2\alpha_8-\alpha_5)$ and $\alpha_5$ are
 somewhat larger than the results of UKQCD \cite{UKQCD:CHPT}:
 $(2\alpha_8-\alpha_5)=0.36 \pm 0.10$ and 
 $\alpha_5=1.22 \pm 0.11$ (only statistical errors quoted).

 The double ratio of the pion mass squares \cite{JLQCD:CHLOGM}
\begin{eqnarray}
RRm \hspace*{-1.5em}&& \equiv \frac{m_{VS}^4}{m_{VV}^2 m_{SS}^2}
= \frac{(\xi+1)(\xi^2+\xi-\eta+2\eta\xi-\eta\xi^2)}{4\xi^2}
\nonumber
\\[0.5em] &&
+ \frac{\chi_S(\xi+1)(\xi^2+\xi+\eta+3\eta\xi^2)\log(\xi+\eta)}
{64N_s\pi^2\xi^2}
- \frac{\chi_S(\xi+1)^2(2\eta+1)\log(1+\eta)}{64N_s\pi^2\xi}
\nonumber
\\[0.5em] && \label{eq26}
- \frac{\chi_S(\xi-1)(\xi+1)^2(\xi+\eta)}{64N_s\pi^2\xi^2}
+ \frac{2N_s\chi_S\eta(\xi+1)(\xi-1)^2}{\xi^2}(L_{S4}-W_{S6})
\end{eqnarray}
 can be used to determine the fifth parameter $(L_{S4}-W_{S6})$.
 In this case a single parameter fit with fixed $\chi_S = 10.0$ and
 $\eta = 0.10$ gives $(L_{S4}-W_{S6})=0.00358(6)$.

 The conclusion of this paper is that -- once the quark masses are
 small enough -- the qualitative behaviour of the low energy chiral
 effective theory with chiral logarithms is present even on coarse
 lattices.
 Since here ratios of pion masses and decay constants are considered
 the $Z$-factors of renormalization cancel, therefore the uncertainty
 about their $\beta$-dependence, which can influence the results of
 \cite{NF2TEST,PRICE}, is removed.
 The coefficients of the observed chiral logarithms and the fitted
 values of the Gasser-Leutwyler coefficients are close to expectation.
 This qualitative agreement of the results of a numerical simulation
 with (PQ)ChPT is quite satisfactory but for a quantitative
 determination of NLO ChPT parameters one has to perform
 extrapolations to $a \to 0$ and $m_q \to 0$.
 Since ${\cal O}(a)$ effects are taken into account in the analysis
 by the Rupak-Shoresh effective Lagrangian, the continuum limit will
 be reached asymptotically at the rate ${\cal O}(a^2)$.

 The computations were performed on the APEmille systems installed 
 at NIC Zeuthen, the Cray T3E systems at NIC J\"ulich and the PC
 clusters at DESY Hamburg.
 We thank Hartmut Wittig for helpful discussions on various topics
 and for careful reading of the manuscript.
 I.M. profited from enlightening discussions on the Asia-Pacific
 Mini-Workshop on Lattice QCD especially with Shoji Hashimoto,
 Yoichi Iwasaki, Yusuke Namekawa and Steve Sharpe.



\end{document}